\documentstyle[preprint,aps]{revtex}
\tightenlines
\begin{document}
\draft
\title{Atomic versus molecular diffraction: Influence of
  break-ups and finite size}
\author{Gerhard C. Hegerfeldt}
\address{Institut f\"ur Theoretische Physik,
  Universit\"at G\"ottingen, Bunsenstr. 9,
  D-37073 G\"ottingen, Germany}
\author{Thorsten K\"ohler\cite{koehler}}
\address{Max-Planck-Institut f\"ur Str\"omungsforschung,
  Bunsenstr. 10, D-37073 G\"ottingen, Germany}

\maketitle
\begin{abstract}
Atomic diffraction through double slits and transmission gratings is well
described in terms of the associated de Broglie waves and classical wave
optics. However, for weakly bound and relatively large systems, such as
the He$_2$ dimer, this might no longer hold true due to the possibility of
break-up processes and finite-size effects. We therefore study the
diffraction of weakly bound two-particle systems. If the bar and 
slit widths 
of the grating are much larger than the diameter of the
two-particle system we recover the usual optics results. For smaller
widths, however, deviations therefrom occur. We find that the location of
possible diffraction peaks is indeed still governed by the usual grating
function from optics, but the peaks may have a lower 
intensity. This is not
unexpected when break-up processes are allowed. More unusually though,
diffraction peaks which would be absent for de Broglie waves may reappear. 
The results are illustrated for diffraction of He$_2$. 
\end{abstract}
\section{Introduction}
The rapidly developing new field of atom optics 
\cite{AtOptReviews}
exploits wave
aspects of quantum mechanical particles and associated interference
effects, similar to electron and neutron interferometry. Typical
optics experiments have been carried over to atoms. For example,
an atomic beam can be passed through a double slit or through a
transmission grating and the diffraction pattern can be observed
\cite{firstExpMlynekPrichard}. 
In such  experiments the incoming atoms usually can be
considered as point particles and described by plane waves. The
standard wave-theoretical methods and approximations  of classical optics
(Huygens, Kirchhoff) have been successfully applied in atom optics and
yield in general good agreement with experiment. Indeed, in view of
the success of these methods one might be inclined to think that all
theory needed for atom optics is contained in a good optics book.

New problems and challenges, however, might arise for diffraction
experiments with molecules, in particular with weakly bound
systems. A system, when observed at a nonzero diffraction angle from
the incident direction, clearly has received  a momentum transfer from
the 
grating. For weakly bound systems  this might induce
break-up processes which in turn might change the diffraction
pattern. 
Moreover, weakly bound systems in general are larger than
atoms. Treatment as a point particle might therefore not be adequate
when the grating slits approach the size of the system and when
therefore more break-up processes are to be 
expected. In fact, it is an interesting and
difficult dynamical problem how a relatively large molecule squeezes
through a slit.

In view of recent diffraction experiments with molecules this is not a
moot question. Grating 
diffraction of both He$_2$ \cite{SchoellToennScience} 
and of Na$_2$
\cite{PrichardMolecule} have been observed \cite{Estermann}. 
The molecule Na$_2$ is very small
compared to present-day slit widths and the binding energy is large
compared to the incident molecular kinetic energy. Therefore no
appreciable deviations from point particle diffraction are expected
even if present-day slit widths 
are reduced by a factor of 2 or 4. On the other
hand, for the helium dimer He$_2$, discovered a few years ago
\cite{Luofirstdimer,SchoellToennScience},
the extremely low
binding energy of $-0.11$ $\mu$eV
(see e.g. Ref. \cite{TangToenniesYiu}) 
is much smaller than the incident
kinetic energy of typical experiments. Its estimated diameter of about
6 nm \cite{LuoGieseGentry} is still an order of magnitude 
smaller than the slit widths 
used in the experiments, but a further reduction in the 
slit width might make a difference.

A theoretical analysis of diffraction of weakly bound systems by a
transmission grating, as undertaken in this paper, may therefore be of
more than only academic interest. We restrict ourselves to diatomic
systems. Since diffraction by a transmission grating is really
a  scattering by the bars of the grating and since we want to include 
break-up 
processes and finite-size effects we use quantum mechanical
scattering theory. As in the case of point-particle
diffraction we describe the effect of the grating bars by a 
short-range 
repulsive (``reflecting'') potential. Attractive parts to the
potential could also be included, but for simplicity this is not done
here. 

As usual we neglect the electronic degrees of freedom of the two
atoms. We thus consider a bound system of two interacting
particles and its scattering by an external potential. Due to the
presence of the latter this problem does not resemble so much 
one-particle scattering but rather the three-body problem (the external
potential playing the role of the third body). Not surprisingly one
can therefore use similar methods. We have found it advantageous to
employ the Faddeev approach \cite{Fad} in its formulation by
Alt-Grassberger-Sandhas (AGS) \cite{AGS}.

Briefly, our results are the following. We obtain an analytic
expression for the elastic diffraction amplitude of a weakly bound
two-particle system which -- implicitly -- incorporates break-up
processes and finite-size effects. We also give an expression for
diffraction including excitation and de-excitation  of the two-particle
system, but the break-up channel is not explicitly treated here. 

Provided the bar and slit width of the 
grating are not much smaller than the molecular
diameter, the diffraction amplitude contains  the familiar
grating function of classical optics as a factor. 
This grating function appears
automatically from our general results, and we give a
simple intuitive argument why its appearance is not surprising. The
grating function, which just as in classical optics determines the peaks at
the different diffraction orders, is multiplied by the molecular
diffraction amplitude due to a single bar of the grating, and this is
where a difference to the point-particle case may arise. 

We show that for a molecular diameter very small compared 
to the bar and slit width the
deviations from the point-particle result are indeed negligible, as
expected. This then verifies the interpretation of the He$_2$
diffraction  curves given in Refs. 
\cite{SchoellToennScience,SchoellToennJCP}. 
For smaller bar and 
slit widths noticeable deviations occur. In particular,
heights of diffraction peaks may become lower than for a point
particle. This can clearly be attributed to break-ups which diminish
the actual number of molecules in a given direction. Moreover, the
decrease in peak height goes with the diffraction angle and thus with
the lateral momentum transfer. This is also intuitively clear since
higher momentum transfer implies more break-ups. 
As another effect, diffraction orders 
which are sometimes suppressed to zero 
for point particles may reappear. This is probably a 
finite-size effect.

If the slits become much smaller than the diameter of the two-particle
system, additional terms appear in the diffraction pattern. This case,
however, is presently not of experimental interest and not further
evaluated here.

The plan of the paper is as follows. In Section 
\ref{sec:generaltheory}
we establish AGS
equations for the scattering amplitude of a bound two-particle system
for scattering by an external potential. These equations decouple in our
case. They can be iterated in a straightforward way, and in Section 
\ref{sec:specialization}
we specialize the lowest order term  to diffraction
scattering. In Section 
\ref{sec:evaluation} 
we evaluate this for the diffraction of a
weakly bound two-particle system by a transmission grating. The reader
not interested in the detailed derivation can proceed directly to the main
result in Eqs. (\ref{38}) to (\ref{40}). As an
illustration of our general formulae, in Section 
\ref{sec:example}
diffraction patterns for He$_2$ are determined  for three
slit widths and compared with the corresponding patterns for a point
particle. A full discussion of the He$_2$ case and a detailed
comparison with corresponding 
experimental results will appear elsewhere \cite{HeKoeToe}.

\section{General theory}
\label{sec:generaltheory}
We consider a bound two-particle system -- designated as (diatomic)
``molecule'' -- which is scattered by an obstacle. Later we will
specialize to diffraction by a transmission grating. The two particles
are taken to represent two atoms whose electronic degrees of freedom
are neglected. The (molecular) binding potential of the two particles
is denoted by $V$ which may have several bound states, with (negative)
binding energy $E_\gamma$ and wave function $\phi_\gamma ({\bf x})$
where ${\bf x} \equiv{\bf x}_1 - {\bf x}_2$ and ${\bf x}_1$ 
and ${\bf x}_2$ 
are the coordinates of the two particles. The obstacle
(grating) is represented by a potential $W$ of the form
\begin{equation}
  \label{a}
  W({\bf x}_1,{\bf x}_2)=W_1({\bf x}_1)+W_2({\bf x}_2)
\end{equation}
where $W_1$ and $W_2$ are short-range and strongly repulsive
\cite{Bem0}.

For a molecule, as opposed to a point particle, there are various
scattering channels. In the elastic channel the molecule is scattered
without energy transfer, while in the inelastic channels the molecule
might end up in a different bound state or might break up. In a
break-up process both constituents may emerge as asymptotically free
particles or one of them may stick to the obstacle if $W$ has an
attractive part.

The total Hamiltonian for the problem is $H_0 + V + W$, with $H_0$ the
kinetic energy,
\begin{equation}
  \label{1}
  H_0\equiv\hat{{\bf p}}_1^2/2m_1+\hat{{\bf p}}_2^2/2m_2
  =\hat{{\bf P}}^2/2M+\hat{{\bf p}}^2/2\mu
\end{equation}
where $\hat{{\bf p}}_1$ and $\hat{{\bf p}}_2$ are the momentum
operators of the particles, $\hat{{\bf P}}$ and $\hat{{\bf p}}$ those
for the total and relative momentum, and $M$ and $\mu$ the total and
reduced mass, respectively. For elastic and (de-)excitation
scattering the channel Hamiltonian is $H_0 + V$. Its improper
eigenstates are denoted by $|{\bf P}, \phi_\gamma \rangle$,
\begin{equation}
  \label{2}
  (H_0+V)|{\bf P},\phi_\gamma\rangle=({\bf P}^2/2M+E_\gamma)
  |{\bf P},\phi_\gamma\rangle \equiv E |{\bf P},\phi_\gamma\rangle.
\end{equation}
As in the case of one-particle scattering the $S$ matrix can be
decomposed as 
\begin{equation}
  \label{3}
  \langle{\bf P},\phi_\gamma|S|{\bf P}^\prime,\phi_{\gamma^\prime}
  \rangle=\delta_{\gamma\gamma^\prime}\delta^{(3)}
  ({\bf P}-{\bf P}^\prime)
  -2\pi{\rm i}\delta(E-E')
  t({\bf P},\phi_\gamma;{\bf P}^\prime,\phi_{\gamma^\prime})
\end{equation}
where $t({\bf P}, \phi_\gamma; {\bf P}', \phi_{\gamma^\prime})$ 
is the
transition amplitude for an incoming molecule of momentum ${\bf P}'$
and internal state $\gamma'$ to an outgoing one of momentum ${\bf P}$
and internal state $\gamma$.

In one-particle scattering the scattering amplitude can usually be
determined by means of a Lippmann-Schwinger (LS) equation. In the
present multi-channel situation there is a set of coupled LS
equations, and one meets the same difficulties with them as in the
three-body problem. In the latter one therefore uses the Faddeev
approach \cite{Fad}. For the calculation of transition amplitudes one
often employs its formulation by 
Alt-Grassberger-Sandhas (AGS) \cite{AGS}. It
turns out that the AGS approach is easily carried over to the present
two-body problem with its external potential. Moreover, the resulting
equations decouple and can be evaluated for interesting
situations. This will now be outlined. 

We introduce the Green's operators
\begin{eqnarray}
  \label{6}
  G_0(z)&\equiv&(z-H_0)^{-1}\\
  \label{7}
  G(z)&\equiv&(z-H_0-V-W)^{-1}\\
  \label{8}
  G_V(z)&\equiv&(z-H_0-V)^{-1}
\end{eqnarray}
and similarly $G_W(z)$. Here $G_0$ is the free and $G$ the total
resolvent. One has the usual resolvent equations
\begin{eqnarray}
  \label{c}
  &G=G_V+G_V WG\\
  \label{d}
  &G=G_W+G_W VG.
\end{eqnarray}
We also define
\begin{eqnarray}
  \label{5}
  T_V(z)&\equiv&V+VG_V(z)V\\
  \nonumber
  T_W(z)&\equiv&W+WG_W(z)W
\end{eqnarray}
which are the $T$ matrices \cite{Gloeckle} for 
the potential $V$ and $W$, respectively, in the
two-particle Hilbert space. They satisfy
\begin{eqnarray}
  \label{17}
  &T_VG_0=VG_V\\
  \nonumber
  &T_WG_0=WG_W
\end{eqnarray}
which follows by writing Eq. (\ref{c}) as
\begin{equation}
  T_V=VG_V(G_V^{-1}+V)=VG_VG_0^{-1}
\end{equation}
and similarly for $T_W$. 

In analogy to the treatment of the three-body case in Ref. \cite{5}
 we now
introduce operators $U_{VV}(z)$ and $U_{WV}(z)$ through the equations
\begin{eqnarray}
  \label{e}
  G(z)&=&G_V(z)+G_V(z)U_{VV}(z)G_V(z)\\
  \label{f}
  G(z)&=&\ \ \ \ \ \ \ \ \ \ \ \ G_W(z)U_{WV}(z)G_V(z).
\end{eqnarray}
Proceeding verbatim as in Section 7 of Ref. \cite{5}
one shows that the
transition amplitudes of Eq. (\ref{3}) are given by
\begin{equation}
  \label{12}
  t({\bf P},\phi_\gamma;{\bf P}^\prime,\phi_{\gamma^\prime})=
  \langle{\bf P},\phi_\gamma|U_{VV}(E+{\rm i}0)
  |{\bf P}^\prime,\phi_{\gamma^\prime}\rangle
\end{equation}
where
\begin{equation}
  \label{g}
  E={\bf P}^{\prime 2}/2M+E_{\gamma^\prime}=
  {\bf P}^2/2M+E_\gamma.
\end{equation}
Thus $U_{VV}$ acts as a transition operator, and its advantage is that it
satisfies a simple equation, seen as follows. We equate Eqs. (\ref{c})
and (\ref{e}), 
\begin{equation}
  G_V+G_VWG=G_V+G_VU_{VV}G_V,
\end{equation}
and insert Eq. (\ref{f}) for $G$. This gives 
\begin{equation}
  \label{h}
  U_{VV}=WG_WU_{WV}=T_WG_0U_{WV}
\end{equation}
by Eq. (\ref{17}). Similarly equating Eqs. (\ref{d}) and (\ref{f}) and
inserting Eq. (\ref{e}) gives
\begin{equation}
  \label{i}
  U_{WV}=G_0^{-1}+T_VG_0U_{VV}.
\end{equation}
Eqs. (\ref{h}) and (\ref{i}) are the AGS equations for our problem
which decouple here. 
Indeed, insertion of
Eq. (\ref{i}) into Eq. (\ref{h}) gives 
\begin{equation}
  \label{18}
  U_{VV}=T_W+T_WG_0T_VG_0U_{VV}
\end{equation}
which no longer contains $U_{WV}$. This is an exact equation for the
transition operator and for the transition amplitudes in
Eq. (\ref{12}). Break-up processes and finite-size effects
are exactly included. 

Similar to Eq. (\ref{12}) one
can show that 
\begin{equation}
  \label{19}
  U_{0V}\equiv G_0^{-1}+T_VG_0U_{VV}+T_WG_0U_{WV}
\end{equation}
is the transition operator for break-up processes into two
asymptotically free particles. Thus a knowledge of $U_{VV}$ and
$U_{WV}$ would also cover these processes. One can further show that
Eqs. (\ref{h}) and (\ref{i}) yield unitarity of the $S$ matrix.

The great advantage of Eq. (\ref{18}) or of the system in (\ref{h})
and (\ref{i}) is that it lends itself immediately to a perturbative
approach by iteration. Such an iterative procedure is also used in the
multiple-scattering expansion in the three-body problem 
\cite{Gloeckle}. To
lowest order one has
\begin{equation}
  U_{VV}\cong T_W
\end{equation}
and as in Ref.
\cite{Gloeckle}
this is expected to give a good approximation for large
incident energy, viz. for 
\begin{equation}
  \label{20}
  P^{\prime 2}/2M\gg |E_g|,\langle \phi_g|V|\phi_g\rangle
\end{equation}
where $P' = |{\bf P}'|$, $E_g$ is 
the ground-state energy and $\phi_g$
the ground-state wave-function. Then in general the above condition
also holds for other bound states. The relevant 
matrix elements of $T_V$ will
then be small compared to the incident energy and this allows one to
show that the next order term in the iteration can be
neglected. Thus in the region of large incident energy
one has, by Eq. (\ref{12}) for the transition amplitude in
Eq. (\ref{3}),
\begin{equation}
  \label{21}
  t({\bf P},\phi_\gamma ;{\bf P}^\prime,\phi_{\gamma^\prime})
  \cong \langle {\bf P},\phi_\gamma |T_W(E+{\rm i}0)
  |{\bf P}^\prime,\phi_{\gamma^\prime}\rangle.
\end{equation}

Although $T_W$ is the transition operator for scattering of two 
asymptotically free
particles by $W$, the above matrix element differs crucially from a
scattering  amplitude for free particles since the wave functions
$\phi_\gamma$ and $\phi_{\gamma'}$ do not have a definite (relative)
momentum. It is therefore through these wave functions that the above
expression takes the interaction potential $V$ between the two
particles and the finite size of the molecule into account.

The major problem remaining is the explicit evaluation of Eq. (\ref{21})
for the transition amplitude, and this we will do here for the
experimentally interesting case of diffraction by a grating and for
small scattering angles. 

\section{Specialization to diffraction scattering}
\label{sec:specialization}
The operator $T_W(z)$ appearing in the approximate expression for 
the transition amplitude in Eq. (\ref{21}) is, in principle, the 
$T$ matrix 
for the scattering of two asymptotically
free particles by the potential 
$W=W_1 + W_2$. 
But even for this simple additive potential it is
not easy to calculate $T_W (z)$ since (i) it is not evaluated on the
energy shell of two asymptotically 
free particles and (ii) the Green's operator $G_W$
is not easily expressed by one-particle Green's operators. We
therefore restrict ourselves in the following to the experimentally
important diffraction domain and to small scattering angles.

In the diffraction domain the de Broglie wave-length is,
by definition, small
compared to the extension of the obstacle and the potential
\cite{Sitenko}. Small
scattering angles mean $|{\bf P}_\perp | \ll P'$ where 
${\bf P}_\perp$ is the
component of ${\bf P}$ orthogonal to the incident momentum 
${\bf P}'$. 
For a transmission grating of $N$ bars, grating period $d$ and
slit width $s$ the domain of diffraction scattering can be characterized
by the condition 
\begin{equation}
  \label{23}
  P^\prime s/\hbar,P^\prime (d-s)/\hbar\gg 1.
\end{equation}
We also assume here and in the following that
the masses 
$m_1$ and $m_2$ are of
the same order of magnitude.

To calculate the matrix element of $T_W$ in Eq. (\ref{21}) one could
proceed as in the eikonal approximation \cite{Sitenko} by considering
an absorbing (``black'') obstacle, instead of a reflecting one, and
thus work with a complex {\em ersatz} potential instead of the real
potential $W$. We have done this, and the results agree
with those of the more direct approach we are presenting here. 

Inserting two complete sets of two-particle plane waves in the
r.h.s. of Eq. (\ref{21}) and performing the integration over resulting
$\delta$ functions one obtains
\begin{eqnarray}
  \label{24}
  &\langle {\bf P},\phi_\gamma|T_W(E+{\rm i} 0)
  |{\bf P}^\prime ,\phi_{\gamma^\prime}\rangle
  =\int {\rm d}^3 p {\rm d}^3 p' \phi_\gamma^*({\bf p})\\
  \nonumber
  &\times\langle {\bf p}_1,{\bf p}_2|
  T_W(E+{\rm i} 0)|{\bf p}_1^\prime,{\bf p}_2^\prime\rangle 
  \phi_{\gamma^\prime}({\bf p}^\prime)
\end{eqnarray}
where for ${\bf p}_1$ and ${\bf p}_2$ one has to insert
\begin{equation}
  \label{25}
  {\bf p}_{1,2}=\frac{m_{1,2}}{M}{\bf P}\pm {\bf p}
\end{equation}
and similarly for ${\bf p}'_{1,2}$. We note that the kinetic energies
of $|{\bf p}_1 , {\bf p}_2 \rangle$ and 
$| {\bf p}'_1, {\bf p}'_2\rangle$ 
are not equal to $E$ so that we are not on the energy
shell. However, in Eq. (\ref{25}) only those values of ${\bf p}$ and
${\bf p}'$ enter for which $\phi_\gamma ({\bf p})$ and 
$\phi_{\gamma^\prime} ({\bf p}')$ 
are essentially different from zero. The condition Eq.
(\ref{20}) of large incident energy ensures that 
$\phi_\gamma ({\bf  p})$ and $\phi_{\gamma^\prime}({\bf p'})$ 
are centered at momenta which are
small compared to ${P'}$ since the Schr\"odinger equation implies,
upon multiplication with $\langle \phi_\gamma|$, 
\begin{equation}
  \langle\phi_\gamma|\hat{\bf p}^2|\phi_{\gamma}\rangle/2\mu
  =E_\gamma-\langle\phi_\gamma|V|\phi_\gamma\rangle\ll
  P^{\prime 2}/2M.
\end{equation}
Therefore one has indeed 
\begin{equation}
  \label{j_1}
  |{\bf p}|,|{\bf p}^\prime|\ll P^\prime\approx |{\bf P}|\equiv P
\end{equation}
and thus, by Eq. (\ref{25}), 
\begin{eqnarray}
  \label{j_2}
  &p_{1,2}^\prime\equiv
  |{\bf p}_{1,2}^\prime| \approx \frac{m_{1,2}}{M} P^\prime\\
  \nonumber
  &p_{1,2}\equiv|{\bf p}_{1,2}|\approx \frac{m_{1,2}}{M} P.
\end{eqnarray}
Hence
\begin{equation}
  \label{j}
  E_1^\prime+E_2^\prime\equiv 
  {\bf p}_1^{\prime 2}/2m_1+
  {\bf p}_2^{\prime 2}/2m_2
  \approx P^{\prime 2}/2M
  \approx E
\end{equation}
where in the last step Eqs. (\ref{g}) and (\ref{20}) have been
used. Now one can convince oneself
by explicit calculation that under the above
conditions
$\langle{\bf p}_1,{\bf p}_2|T_W(E+{\rm i} 0)
  |{\bf p}_1^\prime,{\bf p}_2^\prime\rangle$
is slowly varying in $E$ so that $E$ can be replaced by the l.h.s. of
Eq. (\ref{j}). Hence
\begin{eqnarray}
  \label{k}
  &\langle{\bf p}_1,{\bf p}_2|T_W(E+{\rm i} 0)
  |{\bf p}_1^\prime,{\bf p}_2^\prime\rangle\\
  \nonumber
  &\cong\langle{\bf p}_1,{\bf p}_2|
  T_W({\bf p}_1^{\prime 2}/2m_1+
  {\bf p}_2^{\prime 2}/2m_2+{\rm i} 0)
  |{\bf p}_1^\prime,{\bf p}_2^\prime\rangle.
\end{eqnarray}
We now use the LS equation for the potential $W$ and the corresponding
two-particle scattering states $|{\bf p}'_1, {\bf p}'_2, + \rangle$,
\begin{equation}
  |{\bf p}'_1, {\bf p}'_2, + \rangle=|{\bf p}'_1, {\bf p}'_2\rangle
  +G_W({\bf p}_1^{\prime 2}/2m_1+
  {\bf p}_2^{\prime 2}/2m_2+{\rm i} 0)W
  |{\bf p}'_1, {\bf p}'_2\rangle.
\end{equation}
Using Eq. (\ref{5}) for $T_W$, pulling out a factor of 
$W$ and then using
the LS equation we obtain 
$\langle{\bf p}_1,{\bf p}_2|W|{\bf p}_1^\prime,
{\bf p}_2^\prime,+\rangle$
for the r.h.s. of Eq. (\ref{k}).
Now, for $W = W_1 + W_2$ the two-particle scattering state is just
the product of the one-particle scattering states 
$| {\bf p}'_1 , +\rangle_1$ and $|{\bf p}'_2, + \rangle_2$ 
for $W_1$ and $W_2$,
respectively, and $|{\bf p}_1, {\bf p}_2 \rangle = |{\bf p}_1 \rangle
| {\bf p}_2 \rangle$. We thus have from Eq. (\ref{k})
\begin{eqnarray}
  \label{n}
  &\langle{\bf p}_1,{\bf p}_2|T_W(E+{\rm i} 0)
  |{\bf p}_1^\prime,{\bf p}_2^\prime\rangle\\
  \nonumber
  &\cong
  \langle{\bf p}_2|\langle{\bf p}_1|W_1+W_2
  |{\bf p}_1^\prime,+\rangle_1
  |{\bf p}_2^\prime,+\rangle_2.
\end{eqnarray}

To evaluate this we introduce one-particle amplitudes
\begin{equation}
  t_1 ({\bf p}_1, {\bf p}'_1)\equiv 
  \langle{\bf p}_1|W_1|{\bf p}_1^\prime,+\rangle_1
\end{equation}
and similarly for $t_2 ({\bf p}_2, {\bf p}'_2)$. For the special case
$p_i^2 = p_i^{\prime 2}$ these amplitudes would just be physical
transition amplitudes \cite{Newton}, but in general they are off the
energy shell. We also need 
$\langle {\bf p}_1 | {\bf p}'_1, +\rangle_1$, 
and for this we use the LS equation
\begin{equation}
  \label{32}
  |{\bf p}_1^\prime,+\rangle_1=|{\bf p}_1^\prime\rangle
  +(E_1^\prime-\hat{\bf p}_1^2/2m_1+{\rm i}0)^{-1}W_1|
  {\bf p}_1^\prime,+\rangle_1
\end{equation}
where $E'_1 = p^{\prime 2}_1/2 m_1$ 
in the free one-particle Green's
operator. One obtains 
\begin{equation}
  \label{p}
  \langle{\bf p}_1|{\bf p}_1^\prime,+\rangle_1=
  \delta^{(3)}({\bf p}_1-{\bf p}_1^\prime)
  +\frac{t_1({\bf p}_1,{\bf p}_1^\prime)}{E_1^\prime-E_1+{\rm i}0}
\end{equation}
and similarly for particle 2. In this way the r.h.s. of Eq. (\ref{n})
can be written in terms of the one-particle amplitudes $t_1$ and $t_2$
and inserted into Eq. (\ref{24}). We have thus completely expressed
the transition amplitude in Eq. (\ref{21}) in terms of one-particle
amplitudes. 

To further evaluate the integrals over ${\rm d}^3 p$ and 
${\rm d}^3 p'$ in
Eq. (\ref{24}) we use the condition of large incident momentum and
small scattering angles. As outlined above this makes ${\bf p}$ small
in Eq. (\ref{25}), and correspondingly for ${\bf p}'$. A simple
calculation then shows that
\begin{equation}
  \label{q}
  \frac{1}{E_1^\prime-E_1+{\rm i}0}+\frac{1}{E_2^\prime-E_2+{\rm i}0}
  \approx-\frac{2\pi {\rm i}M}{P}
  \delta(p_\parallel-p_\parallel^\prime)
\end{equation}
where $\parallel$ 
denotes the component in the incident direction,
i.e. parallel to ${ \bf P}'$. In order to make Eq. (\ref{q})
applicable we use that $t_i ({\bf p}_i, {\bf p}'_i)$
depends only weakly on $p_\parallel$ and  
$p'_{\|}$ 
since in Eq. (\ref{25})
for ${\bf p}_i$ and ${\bf p}^\prime_i$
one has $|{\bf p}^\prime|,|{\bf p}|\ll P^\prime,P$, by 
Eq. (\ref{j_1}) \cite{Bem4}. 
With Eq. (\ref{q}) an elementary
calculation then gives for the transition amplitude
\begin{eqnarray}
  \label{33}
  & t({\bf P},\phi_\gamma;
  {\bf P}^\prime,\phi_{\gamma^\prime})\cong
  \int{\rm d}^3 p^\prime\phi_\gamma^*
  ({\bf p}^\prime-\frac{m_1}{M}({\bf P}-{\bf P}^\prime))\\
  \nonumber
  & \times t_2({\bf P}-{\bf P}^\prime+\frac{m_2}{M}{\bf P}^\prime
  -{\bf p}^\prime,\frac{m_2}{M}{\bf P}^\prime-{\bf p}^\prime)
  \phi_{\gamma^\prime}({\bf p}^\prime)\\
  \nonumber
  &+\int{\rm d}^3 p^\prime\phi_\gamma^*
  ({\bf p}^\prime+\frac{m_2}{M}({\bf P}-{\bf P}^\prime))\\
  \nonumber
  & \times t_1({\bf P}-{\bf P}^\prime+\frac{m_1}{M}{\bf P}^\prime
  +{\bf p}^\prime,\frac{m_1}{M}{\bf P}^\prime+{\bf p}^\prime)
  \phi_{\gamma^\prime}({\bf p}^\prime)\\
  \nonumber
  & -\frac{2\pi {\rm i} M}{P}\int{\rm d}^2 p_\perp\int{\rm d}^3 p'
  \phi_\gamma^*(p_\parallel^\prime ,{\bf p}_\perp)
  t_1(\frac{m_1}{M}P_\parallel+p_\parallel^\prime,\frac{m_1}{M}
  {\bf P}_{\perp}+{\bf p}_\perp;
  \frac{m_1}{M}{\bf P}^\prime+{\bf p}^\prime)\\
  \nonumber
  & \times
  t_2(\frac{m_2}{M}P_\parallel-p_\parallel^\prime,
  \frac{m_2}{M}{\bf P}_{\perp}-{\bf p}_\perp;
  \frac{m_2}{M}{\bf P}^\prime-{\bf p}^\prime)
  \phi_{\gamma^\prime}({\bf p}^\prime).
\end{eqnarray}
This result holds in the diffraction domain, for small scattering
angles and for $m_1$ and $m_2$ of the same order of magnitude.

\section{Evaluation for diffraction gratings}
\label{sec:evaluation}
We now consider small-angle diffraction of a weakly bound two-particle
system (molecule) by a transmission grating in normal incidence (see
Fig. \ref{Fig1}).
By $\perp$ we denote the component of a vector orthogonal to
${\bf P}'$ and by $\parallel$ 
the component parallel to ${\bf P}'$. Since the
diffraction condition and $|{\bf p}' | , |{\bf p}| \ll P^\prime,P$ 
hold, by
Eq. (\ref{j_1}), the one-particle amplitudes appearing 
in Eq. (\ref{33}) are
only weakly dependent on $p_{\|}$ and $p^\prime_{\|}$. 
One can therefore take them on-shell. Under our
conditions on ${\bf p}_i$ and ${\bf p}^\prime_i$ 
the one-particle amplitudes
are then of the form
\begin{equation}
  \label{r}
  t_i({\bf p}_i,{\bf p}_i^\prime)\equiv
  t(p_i/m_i;{\bf p}_{i\perp}-{\bf p}_{i\perp}^\prime).
\end{equation}
An explicit expression will be given later, but the dependence on the
lateral momentum transfer already allows a simplification of
Eq. (\ref{33}). Using $p_{i}/m_i\approx P/M$, by
Eqs. (\ref{j_1}) and (\ref{25}), one obtains in a straightforward
way
\begin{eqnarray}
  \label{s}
  &t({\bf P},\phi_\gamma;{\bf P}^\prime,\phi_{\gamma^\prime})=
  t(P/M;{\bf P}_\perp)
  \left\{
    F_{\gamma\gamma^\prime}(\frac{m_2}{M}{\bf P}_\perp)
    +F_{\gamma'\gamma}^*(\frac{m_1}{M}{\bf P}_\perp)
  \right\}\\
  \nonumber
  &-\frac{2\pi {\rm i}M}{P}\int{\rm d}^2 q \
  t(P/M;\frac{m_1}{M}{\bf P}_\perp+{\bf q})
  t(P/M;\frac{m_2}{M}{\bf P}_\perp-{\bf q})
  F_{\gamma\gamma^\prime}({\bf q})
\end{eqnarray}
where
\begin{equation}
  \label{t}
  F_{\gamma\gamma^\prime}({\bf q})\equiv
  \int{\rm d}^3 x \phi_\gamma^*({\bf x}){\rm e}^{{\rm i}{\bf q}
    \cdot {\bf x}/\hbar}\phi_{\gamma^\prime}({\bf x}).
\end{equation}
For $\gamma = \gamma'$ the latter
is the Fourier transform of $| \phi_\gamma
({\bf x})|^2$. $F_{\gamma \gamma} ({\bf q})$ is even and real, 
and it could be
called a molecular form factor. Again we recall that we have
assumed diffraction scattering, small angles, and $m_1$ and $m_2$ of
the same order of magnitude
\cite{Deuteron}.
 
From now on we consider {\em elastic} scattering 
$(\gamma =\gamma')$. 
For a transmission grating of $N$ bars as in Fig. \ref{Fig1} the
one-particle amplitude in Eq. (\ref{r}) is calculated
for reflecting bars by standard
methods \cite{MorseF,bar} as 
\begin{equation}
  \label{u}
  t(p_i/m_i;{\bf q}_\perp)=H(q_2)t_{\rm bar}^{\rm PP}
  (p_i/m_i;q_2)\delta(q_3)
\end{equation}
where ${\bf q}_\perp \equiv {\bf p}_{i \perp} - {\bf p}'_{i \perp}$,
$H$ is the usual grating function
\cite{BornWolf}
\begin{equation}
  \label{35}
  H(q_2)=\frac{\sin(q_2 N d/2\hbar)}{\sin(q_2 d/2\hbar)}
\end{equation}
and where $t^{\rm PP}_{{\rm bar}}$ is the point-particle amplitude 
\cite{Bem3}
for a single
bar (in dimension two),
\begin{equation}
  \label{v}
  t_{\rm bar}^{\rm PP}(p_i/m_i;q_2)=
  -\frac{2{\rm i} p_{i}}{(2\pi)^2 m_i}
  \frac{\sin[q_2(d-s)/2\hbar]}{q_2}.
\end{equation}  
The $\delta$ function appears here because we have taken the bars
from $- \infty$ to $\infty$ in the third direction. 
As a consequence of this
the molecular transition amplitude in Eq. (\ref{s}) is a product of
$\delta (P_3)$ and of a smooth function, and the norm-square of the
latter is proportional to the experimentally relevant 
differential cross section.

To further evaluate the transition amplitude we first consider the
simplest case where both the bar and slit width are much larger than
the diameter of the two-particle system. Then the one-particle
amplitude $t(P/M; {\bf P}_\perp)$ entering Eq. (\ref{s}) is sharply
peaked around ${\bf P}_\perp = 0$, on a scale for which 
$F_{\gamma\gamma}(\frac{m_i}{M} {\bf P}_\perp)$ 
will in general vary only very
little and hence one can replace it by $F_{\gamma \gamma} (0) = 1$. A
simple calculation then shows that the integral in Eq. (\ref{s})
cancels half of the first term. The net result is $t(P/M ; {\bf
  P}_\perp)$. This is just the result for a point particle of mass 
$M= m_1+m_2$ and incident and final momentum ${\bf P}'$ and ${\bf P}$,
respectively, yielding the diffraction pattern of classical optics for
wave numbers ${\bf k}' = {\bf P}'/\hbar$ and 
${\bf k} = {\bf P}/\hbar$. 
Thus in the case of bar and slit width much larger than
the diameter of the two-particle system one recovers for the elastic
case the expected point-particle result \cite{Bem2}. 

The situation changes when bar and slit width decrease. The grating
function $H$ in Eq. (\ref{35}) can be written in an elementary way as
a geometric sum of $\exp\{-{\rm i} q_2 d/\hbar \}$, up to a phase 
factor. 
Inserting this into
Eq. (\ref{u}) and subsequently into Eq. (\ref{s}) then yields for the
molecular transition amplitude a simple sum plus a double sum from the
integral term. Combining the diagonal part of the double sum with the
first sum gives what we call the coherent part $t_{{\rm coh}}$ of the
amplitude, while the nondiagonal remainder of the double sum gives the
incoherent part,
\begin{equation}
  \label{38}
  t({\bf P},\phi_\gamma;{\bf P}^\prime,\phi_\gamma)
  =t_{\rm coh}({\bf P}_\perp)+t_{\rm incoh}({\bf P}_\perp).
\end{equation}
The coherent part can be interpreted as the contribution from those
processes where the diatomic
molecule interacts with a single bar, while the
incoherent part contains the interaction with different bars. Provided
$d$ and $s$ are not too small, i.e. still comparable to the molecular
diameter, then $t_{{\rm incoh}}$ can be shown to be {\em negligible}
and an
elementary  calculation gives for the coherent part
\begin{equation}
  \label{39}
  t_{\rm coh}({\bf P}_\perp)=t^{\rm mol}_{\rm bar}(\gamma,P/M;P_2)
  H(P_2)\delta(P_3)
\end{equation}
where $H$ is the grating function of Eq. (\ref{35}) and 
$t_{{\rm bar}}^{{\rm mol}}$, 
the transition
amplitude for a single bar of width $d-s$, is given by
\begin{eqnarray}
  \label{40}
  &t^{\rm mol}_{\rm bar}(\gamma,P/M;P_2)
  =t_{\rm bar}^{\rm PP}(P/M;P_2)
  \int{\rm d}^3 x
  \left\{
    {\rm e}^{{\rm i} m_1 P_2 x_2/M \hbar}
    +{\rm e}^{{\rm i} m_2 P_2 x_2/M \hbar}
  \right\}|\phi_\gamma({\bf x})|^2\\
  \nonumber
  &+\frac{2 {\rm i} P}{(2\pi)^2 M}
  \int{\rm d} x_1 {\rm d} x_3 
  \int_0^{d-s} {\rm d} x_2
  |\phi_\gamma({\bf x})|^2\\
  \nonumber
  &\times
  \left\{
    \sin
    \left[
      \frac{P_2}{\hbar}(\frac{d-s}{2}-\frac{m_2}{M}x_2)
    \right]
    +
    \sin
    \left[
      \frac{P_2}{\hbar}(\frac{d-s}{2}-\frac{m_1}{M}x_2)
    \right]
  \right\}/P_2~,
\end{eqnarray}
with $t^{\rm PP}_{\rm bar}$ as in Eq.
(\ref{v})
\cite{bar}.
These results for elastic diffraction by gratings hold for small
diffraction angles, for strongly repulsive short-range bar potentials
(``reflecting'' bars) and for the two constituent masses of comparable
magnitude.
If bar and slit width become much smaller than the diameter
of the two-particle system,
then $t_{{\rm incoh}}$ can no longer be neglected.

The grating function $H$ in Eq. (\ref{39}) 
with its sharp peaks is the same
as for diffraction of point particles
\cite{Bem5}.
However, instead of $t_{{\rm bar}}^{\rm PP}$ one has now 
$t_{{\rm bar}}^{{\rm mol}}$ 
in which the bound-state wave-function of the diatomic molecule
enters, and this may change the height of the peaks. Indeed,
physically one would expect break-up processes to diminish the
number of bound systems scattered into a given direction. This
does indeed happen for certain diffraction peaks, as will be seen
in the example of the next section. However, also the reverse can
happen, as seen by the following simple argument for equal bar and
slit width ($s = d/2$). In this case the point particle amplitude
$t_{{\rm bar}}^{\rm PP}$ for a single bar has a zero at the
even-order
peaks of the grating function and hence these
diffraction peaks vanish for point particles. But in the molecular
case $t_{{\rm bar}}^{{\rm mol}}$ need not be zero there so that
even-order diffraction peaks may reappear. This behavior is not
so easily understood in terms of break-up processes and is probably a
finite-size effect.

\section{Example: Diffraction of helium dimers}
\label{sec:example}
Our result in Eqs. (\ref{39}) to (\ref{40})
for the diffraction of
weakly bound two-particle systems by a transmission grating can be
directly applied to 
the helium dimer ${\rm He}_2$. 
First of all, the binding energy of ${\rm He}_2$ is
very small, $E_b / k_B \approx - 1.3$ mK
(see e.g. Ref. \cite{TangToenniesYiu}). 
There seems to be
only a single bound state and the estimated 
bond length of about $6$ nm
is enormous \cite{LuoGieseGentry}. 
The helium dimer ${\rm He}_2$ is
by far the largest 
of known diatomic molecules when considered in their ground state
\cite{Julienne}. Our assumption of a
totally reflecting one-particle potential seems to be well suited to
He \cite{LuoGieseGentry}.
The bound-state wave-function $\phi_b ({\bf x}) \equiv \phi_b(r)$
is rotationally symmetric ($s$ state). 
For our numerical calculations we employ the 
analytical expression of $\phi_b(r)$
given in Ref.
\cite{RickLynchDoll}. 
Rotational symmetry allows one to simplify the molecular
transition amplitude $t_{{\rm bar}}^{{\rm mol}}$ for a single bar in
Eq. (\ref{40}) and to express the multiple integrals as a sum of
integrals over $r$. In Fig. \ref{Fig2} 
we have evaluated $\left| t_{{\rm bar}}^{{\rm mol}}\right|^2$ 
for ${\rm He}_2$ as a function of the lateral
momentum transfer $P_2$ divided by $\hbar$ (``wave number'' $k_2$) for
various bar widths and have compared it with 
$\left| t_{\rm bar}^{\rm PP}\right|^2$ 
for the corresponding point particle of equal mass as ${\rm He}_2$
(dashed line, from Eq. (\ref{v})).

For a grating period of $d = 100$ nm and slit width $s = 50$ nm 
there
is little difference between $t_{{\rm bar}}^{{\rm mol}}$ 
and $t_{{\rm bar}}^{\rm PP}$ 
for lateral momentum transfer up to the fifth order
diffraction peak of the grating, 
and so the overall diffraction pattern in Fig. \ref{Fig3}a
agrees well with that of a point particle
\cite{Bem2}. 
Since in this case the
diameter of ${\rm He}_2$ is small compared to the 
slit width this verifies
the result of the discussion in the preceding section. For $d = 50$ nm
and $s=25$ nm the difference in Fig. \ref{Fig2}b is more pronounced. First
of all, at the odd-order peaks of the grating function -- which
correspond to the maxima of $| t_{{\rm bar}}^{\rm PP}|^2$, to
excellent approximation -- $|t_{\rm bar}^{{\rm mol}}|^2$ is 
smaller than $|t_{{\rm bar}}^{\rm PP}|^2$ and hence the
resulting diffraction peaks should be smaller. 
This is indeed seen in Fig. \ref{Fig3}b and can be attributed
to break-ups. Second, $t_{{\rm
    bar}}^{{\rm mol}}$ is nonzero at the zeros of $t_{{\rm bar}}^{\rm PP}$
and hence, by the discussion of the preceding section, the even-orders
diffraction peaks should start to reappear. Again this is seen in
Fig. \ref{Fig3}b.  Both phenomena become more and more pronounced for bar and
slit width still closer to the ${\rm He}_2$ diameter. This is seen in
Figs. \ref{Fig2}c and \ref{Fig3}c.

A more detailed study of ${\rm He}_2$ diffraction and comparison with
experimental results is under way 
\cite{HeKoeToe}.

\section{Conclusions}
In atom optics the finite size of the atoms can usually be ignored,
and for instance atom diffraction by a transmission grating is
exceedingly well described by matter waves for a point particle, 
using the formulae of classical wave optics. 
The situation may already change for a weakly bound 
and relatively large molecule of two atoms, such as the helium dimer
He$_2$, which is the
largest presently known diatomic molecule 
(in ground state)
and which is of great current
experimental interest 
\cite{LuoGieseGentry,SchoellToennMolBeams}. 
A molecular diameter which is no
longer tiny compared to the slit width may have an effect on the
diffraction pattern, and the same holds for possible break-ups of a weakly
bound system. 

In this paper we have studied diffraction scattering of a weakly bound
two-particle system, such as He$_2$, and in particular diffraction by
a transmission grating for small diffraction angles. The relevant
formulae for the transition amplitude for elastic diffraction
scattering are given in Eqs. (\ref{38}) to (\ref{40}) 
from which the diffraction
pattern is obtained. Insofar as the same grating function appears, which
determines the allowed diffraction peaks, 
it has a structure similar to that for 
waves which one associates with point particles. 
This grating function is now
multiplied, however, by the transition amplitude for a single bar, and
for a diatomic 
molecule this may differ from that for a point particle. If the
bar and slit widths of the grating are much larger than the molecular
diameter there is practically no difference to point particles. For
smaller bar and slit widths the finite size of the molecule and
possible break-up processes may have noticeable effects. For example,
 as expected in the presence
of break-ups, diffraction peaks may have lower height. 
But, somewhat more unusual, diffraction peaks which are
suppressed for point particles may reappear. This is probably due to
finite-size effects.

Our derivation used techniques from the three-body problem, in
particular the Alt-Grassberger-Sandhas equations, to express the
transition amplitude in terms of matrix elements of a transition
operator. Their evaluation can be performed with or without the
eikonal approximation, and the results for both cases agree.

We intend to carry the quantum mechanical methods employed here over to
weakly bound three-particle systems, such as the helium trimer
He$_3$. This is an interesting system, in particular in connection with
Efimov states \cite{Efimov}, and presently under experimental study
\cite{SchoellToennMolBeams}.
\section*{Acknowledgments}
We would like to thank W. Sandhas, W. Sch\"ollkopf and J. P. Toennies
for stimulating discussions.

\begin{figure}[h]
  \caption{Diffraction grating with grating period $d$ and slit width $s$.}
  \label{Fig1}
\end{figure}
\begin{figure}[h]
  \caption{
    Single-bar transition amplitudes for diatomic
    molecule and point particle: 
    $\left| t_{{\rm bar}}^{{\rm mol}}\right|^2$
    (solid line) and 
    $\left| t_{\rm bar}^{\rm PP}\right|^2$
    (dashed line) for a) $d=100$ nm,
    b) $d=50$ nm, and c) $d=25$ nm ($s=d/2$).
    The difference increases for decreasing bar and slit width.}
  \label{Fig2}
\end{figure}
\begin{figure}[h]
  \caption{
    Calculated 
    helium dimer diffraction pattern (solid line) and 
    respective point-particle pattern (dashed line) for a) $d=100$ nm,
    b) $d=50$ nm, and c) $d=25$ nm ($N=30$, $s=d/2$). 
    For smaller bar and slit width the deviation from the 
    point-particle result is increasingly pronounced
    (see also enlarged insets).
    In b) and c) the
    second-order peak noticeably reappears at about 25/(100 nm)
    and 50/(100 nm), respectively.}
  \label{Fig3}
\end{figure}

\begin{thebibliography}{2}
\bibitem{Bem0}
  $W_{1,2}$ could also have an attractive part, but this is not
  considered here.
\bibitem{Deuteron}
  It is noteworthy that the structure of the 
  r.h.s. of Eq. (\ref{s}) is very
  similar to the deuteron  transition amplitude for elastic diffraction
  scattering by a ``black'' nucleus (see Ref. \cite{Sitenko}, p. 354).
\bibitem{bar}
  There is an irrelevant overall phase factor depending on the
  position of the grating.
\bibitem{Bem3}
  It should be noted that for diffraction of point particles
  the amplitude $t_{\rm bar}^{\rm PP}$ can, without loss of
  generality, be replaced by the usual slit function of Ref.
  \cite{BornWolf}
  which is proportional to $\sin(k_2 s/2)/(k_2 s/2)$ where 
  ${\bf k}={\bf P}/\hbar$. This is seen by using the addition formula
  for the sin in Eq. (\ref{v}) and by noting that for 
  $k_2=\pm2\pi n/d$,
  $n=1,2,\ldots$,
  i.e. for the main maxima of the grating function, one has
  $\sin(k_2d/2)=0$ and $\cos(k_2 d/2)=(-1)^n$.
  We also note that in optics one usually considers gratings which are
  closed at the two ends and not open as ours. This leads to a change
  in the zeroth-order peak.
\end{thebibliography}
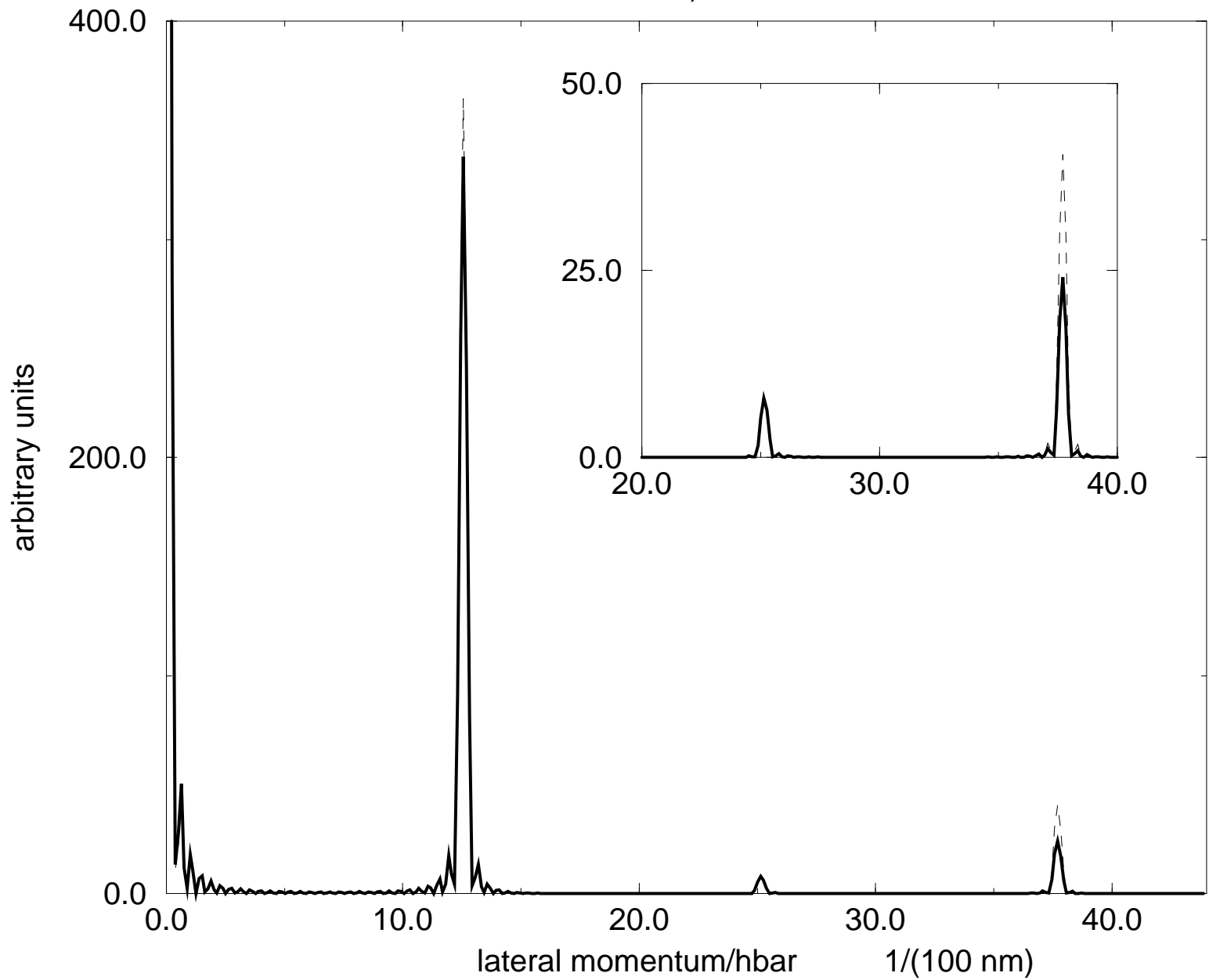
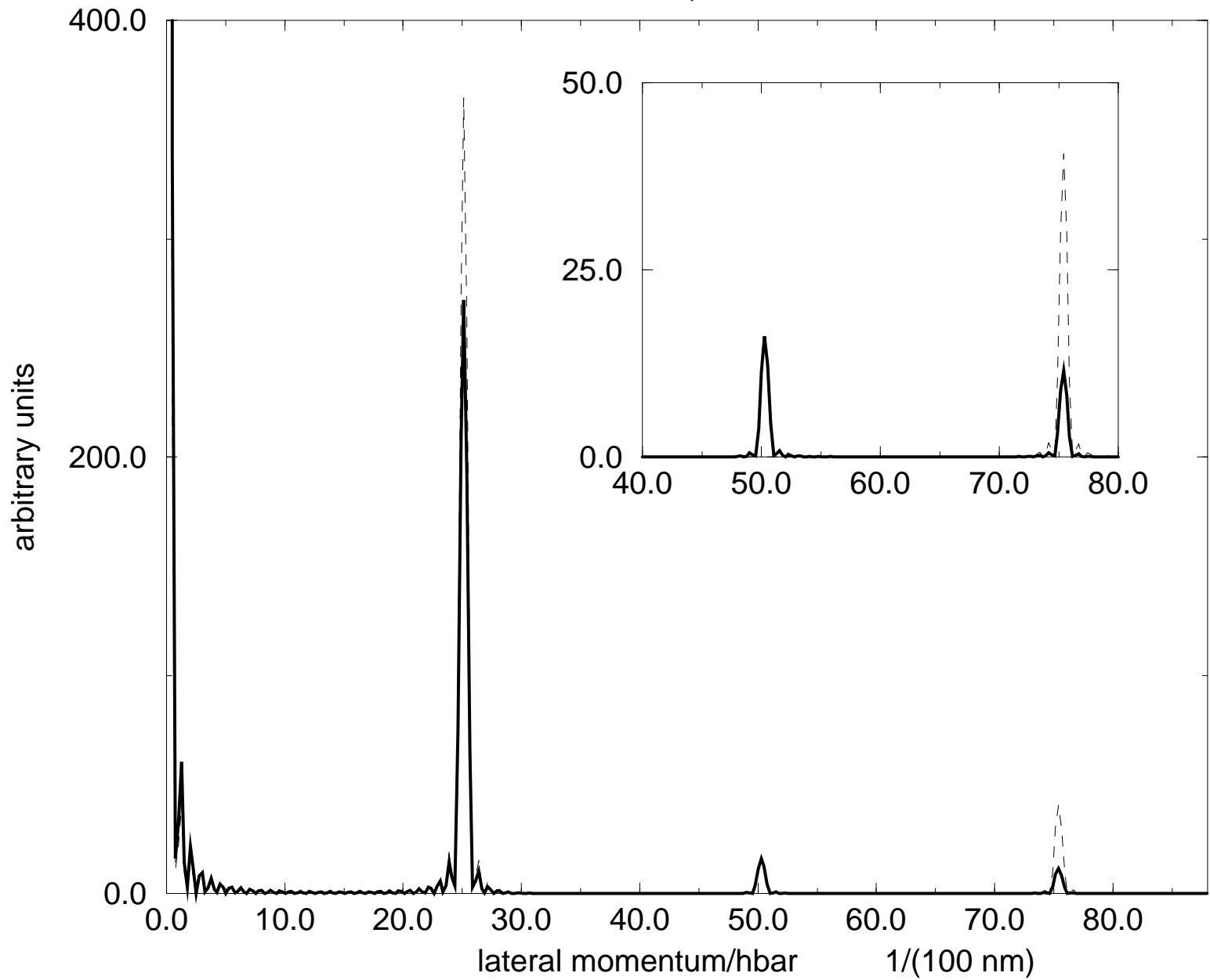
\end{document}